\documentclass[a4paper, 10pt, oneside, onecolumn,3p]{elsarticle}
\journal{Annals of Physics}
\usepackage{amsmath,color}
\usepackage{hyperref}
\usepackage{bm}
\usepackage{mathdots} 
\usepackage{rotating}
\usepackage{amssymb} 
\def\bra#1{\mathinner{\langle{#1}|}}
\def\ket#1{\mathinner{|{#1}\rangle}}
\def\coloneq{\mathrel{\mathop:}=}

\DeclareMathOperator{\dimension}{dim}

\newcommand{\openone}{\mathbf{1}}

\biboptions{sort&compress}

\usepackage[scr=rsfso,scrscaled=1]{mathalfa}

\begin{document}
\title{Heuristic for estimation of multiqubit genuine multipartite entanglement}
\author[uq]{Paulo E. M. F. Mendon\c{c}a\corref{cor1}\fnref{fn1}}
\ead{pmendonca@gmail.com}
\author[jabuca]{Marcelo A. Marchiolli}
\ead{marcelo{\_}march@bol.com.br}
\author[uq]{Gerard J. Milburn}
\ead{milburn@physics.uq.edu.au}
\address[uq]{ARC Centre for Engineered Quantum Systems, School of Mathematics and Physics, The University of Queensland, St.
Lucia, Queensland 4072, Australia}
\address[jabuca]{Avenida General Os\'orio 414, centro, 14.870-100 Jaboticabal, SP, Brazil}
\cortext[cor1]{Corresponding author}
\fntext[fn1]{Permanent address: Academia da For\c{c}a A\'{e}rea, C.P. 970, 13.643-970 Pirassununga, SP, Brazil}
\date{\today}

\begin{abstract}
For every $N$-qubit density matrix written in the computational basis, an associated ``X-density matrix'' can be obtained by vanishing all entries out of the main- and anti-diagonals. It is very simple to compute the genuine multipartite (GM) concurrence of this associated $N$-qubit X-state, which, moreover, lower bounds the GM-concurrence of the original (non-X) state. In this paper, we rely on these facts to introduce and benchmark a heuristic for estimating the GM-concurrence of an arbitrary multiqubit mixed state. By explicitly considering two classes of mixed states, we illustrate that our estimates are usually very close to the standard lower bound on the GM-concurrence, being significantly easier to compute. In addition, while evaluating the performance of our proposed heuristic, we provide the first characterization of GM-entanglement in the steady states of the driven Dicke model at zero temperature.
\end{abstract}

\maketitle

\section{Introduction}

During the last one and a half decades, increasing interest has been manifested in the topic of multipartite entanglement. Just as we learnt, in the early days of quantum information, that \emph{bipartite} entanglement is an invaluable resource for quantum cryptography~\cite{91Ekert661}, communication~\cite{92Bennett2881} and speed-up of classical algorithms~\cite{92Deutsch553,94Simon116,96Grover212,97Shor1484} (to name but a few), we have now a growing awareness of the role that \emph{multipartite} entanglement plays in quantum computing~\cite{03Jozsa2011,11Bruss052313,01Raussendorf5188,09Briegel19}, high-precision metrology~\cite{04Giovannetti1330,12Hyllus022321,13Marchiolli1330001,14Lucke155304,14Vitagliano032307}, quantum phase transitions~\cite{02Osborne032110,05Bruss014301,06Oliveira010305R,06Oliveira039902E1,07Oliveira039901E2,10Montakhab062313,13Giampaolo052305,14Hofmann134101,14Giampaolo93033,14Stasinska032330} and even biology~\cite{10Sarovar462,10Caruso062346}.

Despite significant interest and progresses, many limitations from the theory of bipartite entanglement are naturally inherited in the multipartite setting. For example, since the entanglement detection problem was shown to be NP-hard already in the bipartite case~\cite{03Gurvits10,07Ioannou335}, there is very little hope that computable measures of (multipartite) entanglement can be found for an arbitrary quantum state. As a result, much of the effort in this field is centered around devising \emph{sufficient} criteria for (multipartite) entanglement detection~\cite{96Peres1413,96Horodecki1,00Terhal319,10Huber210501}, which are then turned into computable (multipartite) entanglement estimators~\cite{02Vidal032314,05Brandao022310,11Jungnitsch190502,11Ma062325}. 

Amongst all types of multipartite entanglement, special interest is devoted to the detection and quantification of \emph{genuine multipartite} entanglement, a type of multipartite entanglement established collectively between all $N$ parties of an $N$-partite system. More precisely, a GM-entangled state is any state that cannot be written as a mixture of biseparable states, which, in turn, are those states that are separable with respect to some bipartition of the relevant Hilbert space~\cite{09Guhne1,09Horodecki865}. Besides, it is GM-entanglement that occupies the highest position in the hierarchy of multipartite entanglement~\cite{13Levi150402}.

In this paper, we consider a particular set of sufficient conditions for detecting GM-entanglement introduced by Huber \emph{et al.}~\cite{10Huber210501} --- and turned into an entanglement estimator by Ma \emph{et al.}~\cite{11Ma062325} --- to obtain our main result: a heuristic approach for approximating these estimates with a considerably lower computational cost than that required by the optimization problem appearing in Refs.~\cite{10Huber210501,11Ma062325}. In practice, this result opens the way for systematical and analytical estimation of GM-entanglement in symmetric multiqubit mixed states and, more generally, enables a significant reduction in the computational time required by standard numerical solvers, usually accompanied by some small impact on the accuracy of the estimate.

Our heuristic relies upon three basic facts: (i) the estimates proposed by Ma \emph{et al.}~\cite{11Ma062325} are actually lower bounds on the GM-concurrence --- a GM-entanglement monotone introduced by the authors to generalize the Wootters concurrence~\cite{98Wootters2245} to the multipartite setting; (ii) this lower bound is saturated for the family of $N$-qubit X-states, in which case it is also remarkably easy to compute~\cite{12Rafsanjani062303}; and (iii) in general, it is computationally easier to determine a local-unitary (LU) transformation that approximates a generic density matrix to the X-form, than it is to solve the optimization problem in Refs.~\cite{10Huber210501,11Ma062325}. Accordingly, our so-called \emph{X-heuristic}, estimates the GM-concurrence of an arbitrary $N$-qubit density matrix by applying a LU-transformation that minimizes the entries out of its main- and anti-diagonals and, subsequently, outputting the GM-concurrence of the associated X-density matrix (obtained by neglecting any entry that may have remained out of the main- and anti-diagonals of the LU-transformed density matrix).

In order to illustrate the strengths and drawbacks of the X-heuristic, we explicitly apply it to the family of diagonal symmetric states~\cite{54Dicke99,14Quesada052319,14Wolfe140402} and steady states of the driven Dicke model~\cite{79Puri200,80Drummond1179,81Lawande4171,02Schneider042107}. In both cases, we discuss the trade-off between computational efficiency and accuracy \emph{vis-\`a-vis} the standard scheme for GM-concurrence estimation~\cite{10Huber210501,11Ma062325}. Our comparative analysis reveals the X-heuristic as a useful computational tool for GM-concurrence estimation in multiqubit systems.

As side results, while assessing the X-heuristic with the steady states of the driven Dicke model at zero temperature, we provide the first characterization of multipartite entanglement in this model. In particular, we note that our estimates agree on that GM-entanglement is maximal for the parameter values corresponding roughly to a bifurcation of a fixed point in the corresponding semiclassical dynamics, reinforcing the long-standing (and widely-accepted) conjecture that multipartite entanglement must be maximal at quantum phase transitions~\cite{02Osborne032110}. While the driven Dicke model is a zero-dimensional many-body problem, the bifurcation is controlled by a parameter in the Hamiltonian and the correlations in the quantum steady state change at a particular value of that parameter in analogy to the change in the character of a ground state in a spatially extended many-body quantum phase transition.   Moreover, we show that the two-qubit steady states of this model are examples of non-X states whose concurrence can be \emph{exactly} computed with the simple concurrence formula for X-states, settling an open problem posed in Ref~\cite{12Rafsanjani12043912}.

Our paper is structured as follows. In Sec.~\ref{sec:preliminaries} we review some key concepts relating GM-entanglement and its quantification. In Sec.~\ref{sec:heuristic} our heuristic for GM-concurrence estimation is presented and, in Sec.~\ref{sec:applications}, its accuracy is benchmarked against the standard scheme for the two aforemetioned families of $N$-qubit mixed states. In Sec.~\ref{sec:efficiency} we focus on the computational advantage of our heuristic over the standard scheme. Finally, in Sec.~\ref{sec:conclusion}, we summarize our results and discuss some possible avenues for future work.
\section{Preliminaries}\label{sec:preliminaries}

In this section we briefly review the concept of GM-entanglement and some key results regarding its detection and quantification. In particular, we outline the scheme introduced in Refs.~\cite{10Huber210501,11Ma062325} (here referred to as the ``optimal-$\ket{\Phi}$'' method), a current standard for GM-concurrence estimation in multipartite mixed states. Although we are only interested in multiqubit states, throughout most of this section we leave the dimensionality of each party arbitrary.

\subsection{Genuine Multipartite Entanglement}
An $N$-partite pure state $\ket{\psi}\in \mathcal{H}=\mathcal{H}_1\otimes \mathcal{H}_2\otimes\ldots\otimes \mathcal{H}_N$ (with $\dimension{\mathcal{H}_n}=d_n$), is said to be GM-entangled if it is not biseparable; $\ket{\psi}$ is said to be biseparable, in turn, if there is a bipartition of $\mathcal{H}=\mathcal{H}_A\otimes \mathcal{H}_B$ and a pair of states $\ket{\psi_A}\in \mathcal{H}_A$ and $\ket{\psi_B}\in \mathcal{H}_B$ such that 
\begin{equation}\label{eq:biseparablestructure}
\ket{\psi}=\ket{\psi_A} \otimes \ket{\psi_B}\,.
\end{equation}
In general, there are $\mathfrak{n}\coloneq 2^{N-1}-1$ possible bipartitions of $\mathcal{H}$ and, if a decomposition of the form~(\ref{eq:biseparablestructure}) is found with respect to any one of them, then $\ket{\psi}$ can be immediately declared biseparable. On the other hand, $\ket{\psi}$ can only be declared GM-entangled after ruling out the existence of such a decomposition with respect to every bipartition of $\mathcal{H}$. 

The notions of GM-entanglement and biseparability are extended to mixed states via a convex roof construction: An $N$-partite mixed state $\bm{\rho}$ acting on $\mathcal{H}$ is GM-entangled if it is not biseparable; $\bm{\rho}$ is biseparable if it can be expressed as a convex combination of biseparable pure states, that is,
\begin{equation}\label{eq:cvxsum}
\bm{\rho}=\sum_i p_i\ket{\psi_i}\!\bra{\psi_i}
\end{equation}
where every $\ket{\psi_i}$ is biseparable. Remarkably, although the biseparability of $\bm{\rho}$ requires every $\ket{\psi_i}$ in Eq.~(\ref{eq:cvxsum}) to be biseparable, each one of them can be so with respect to a different bipartition of $\mathcal{H}$, meaning that a biseparable mixed state need not to be separable with respect to any particular bipartition of $\mathcal{H}$.

So far we have only established the notion of GM-entanglement. In what follows we review some recent results on (i) how to tell if a given density matrix $\bm{\rho}$ is biseparable (detection) and (ii) if not, how to estimate the amount of GM-entanglement that it contains (quantification). Needless to say, these are hard problems even in the case $N=2$~\cite{03Gurvits10,07Ioannou335,09Horodecki865}, let alone $N>2$.

\subsection{GM-entanglement detection}
Detection of GM-entanglement has been intensely studied (see, e.g., Ref.~\cite{09Guhne1} for a review). To date, one of the most effective detection schemes was proposed by Huber \emph{et al.}~\cite{10Huber210501}, where it was shown that every biseparable state $\bm{\rho}$ satisfies every inequality of the $\ket{\Phi}$-parametrized family 
\begin{equation}\label{eq:gmecriterion}
\mathscr{I}_{\ket{\Phi}}(\bm{\rho})\leq 0\,,
\end{equation}
where
\begin{equation}\label{eq:Iketphirho}
\mathscr{I}_{\ket{\Phi}}(\bm{\rho})\coloneq \sqrt{\bra{\Phi}\bm{\rho}^{\otimes 2} \bm{\Pi}\ket{\Phi}}-\sum_{i=1}^{\mathfrak{n}}\sqrt{\bra{\Phi}(\bm{\Pi}_{A_i}\otimes \openone_{B_i})^\dagger\bm{\rho}^{\otimes 2}(\bm{\Pi}_{A_i}\otimes \openone_{B_i})\ket{\Phi}}\,.
\end{equation}
In the above, $\ket{\Phi}$ can be chosen as any product state of $\mathcal{H}^{\otimes 2}$, i.e.,
\begin{equation}
\ket{\Phi}=\ket{\mu_1\, \mu_2\, \mu_3\,\cdots\,\mu_N}\otimes \ket{\nu_1\, \nu_2\, \nu_3\,\cdots\,\nu_N}.
\end{equation}
The symbol $\bm{\Pi}$ denotes the global permutation operator that peforms simultaneous permutations of all subsystems across the two copies of $\mathcal{H}$,
\begin{equation}
\bm{\Pi}(\ket{\mu_1\, \mu_2\, \mu_3\,\cdots\,\mu_N}\otimes \ket{\nu_1\, \nu_2\, \nu_3\,\cdots\,\nu_N})=\ket{\nu_1\, \nu_2\, \nu_3\,\cdots\,\nu_N}\otimes\ket{\mu_1\, \mu_2\, \mu_3\,\cdots\,\mu_N}\,,
\end{equation}
whereas $\bm{\Pi}_{A_i}\otimes \openone_{B_i}$ only permutes those subsystems whose labels are in $A_i$ (of a given Hilbert space bipartition $\{A_i|B_i\}$). For example, consider the following bipartitions of the $N$-partite Hilbert space $\mathcal{H}$:
\begin{equation}
\{A_1|B_1\}=\{1|2,3,\ldots,N\}\,,\quad \{A_2|B_2\}=\{2|1,3,\ldots,N\} \quad\mbox{and}\quad \{A_{N+1}|B_{N+1}\}=\{1,2|3,\ldots,N\}\,.
\end{equation}
Then,
\begin{align}
(\bm{\Pi}_{A_1}\otimes\openone_{B_1})\ket{\mu_1\, \mu_2\, \mu_3\,\cdots\,\mu_N}\otimes \ket{\nu_1\, \nu_2\, \nu_3\,\cdots\,\nu_N}&=\ket{\nu_1\, \mu_2\, \mu_3\,\cdots\,\mu_N}\otimes \ket{\mu_1\, \nu_2\, \nu_3\,\cdots\,\nu_N}\,,\\
(\bm{\Pi}_{A_2}\otimes\openone_{B_2})\ket{\mu_1\, \mu_2\, \mu_3\,\cdots\,\mu_N}\otimes \ket{\nu_1\, \nu_2\, \nu_3\,\cdots\,\nu_N}&=\ket{\mu_1\, \nu_2\, \mu_3\,\cdots\,\mu_N}\otimes \ket{\nu_1\, \mu_2\, \nu_3\,\cdots\,\nu_N}\,,\\
(\bm{\Pi}_{A_{N+1}}\otimes\openone_{B_{N+1}})\ket{\mu_1\, \mu_2\, \mu_3\,\cdots\,\mu_N}\otimes \ket{\nu_1\, \nu_2\, \nu_3\,\cdots\,\nu_N}&=\ket{\nu_1\, \nu_2\, \mu_3\,\cdots\,\mu_N}\otimes \ket{\mu_1\, \mu_2\, \nu_3\,\cdots\,\nu_N}\,.
\end{align}
Thus, if one can find a state $\ket{\tilde{\Phi}}$ such that $\mathscr{I}_{\ket{\tilde{\Phi}}}(\bm{\rho})>0$, then $\bm{\rho}$ can be promptly declared to be GM-entangled. Unfortunately, such a criterion does not detect every GM-entanglement [there are GM-entangled density matrices that satisfy the entire family of inequalities~(\ref{eq:gmecriterion})], but it is stronger than many commonly used criteria (see, e.g., \cite{08Seevinck032101} and references therein).

\subsection{GM-concurrence estimation: The optimal-$\ket{\Phi}$ scheme}
The relevance of $\mathscr{I}_{\ket{\Phi}}(\bm{\rho})$ transcends its application in the detection problem, manifesting itself also in the context of GM-entaglement quantification. Indeed, it was  shown by Ma \emph{et al.}~\cite{11Ma062325} that, for every choice of $\ket{\Phi}$, the following inequality holds
\begin{equation}\label{eq:familylb}
C_{GME}(\bm{\rho})\geq \max\left[0, 2\mathscr{I}_{\ket{\Phi}}(\bm{\rho})\right]\,,
\end{equation}
where $C_{GME}(\bm{\rho})$ denotes the \emph{concurrence of GM-entanglement} of $\bm{\rho}$; a well-defined (but difficult to compute) entanglement monotone. 

Unlike $C_{GME}(\bm{\rho})$, the lower bounds $2\mathscr{I}_{\ket{\Phi}}(\bm{\rho})$ are easy to measure and to compute, emerging thus as natural candidates for GM-concurrence estimation. However, the accuracy of such estimates are strongly dependent on the choice of $\ket{\Phi}$ for a given $\bm{\rho}$, highlighting the importance of devising efficient ways of making such a choice.

Formally, the strongest inequality of the family~(\ref{eq:familylb}) is given by
\begin{equation}\label{eq:strongest}
C_{GME}(\bm{\rho})\geq \max\left[0,2\max_{\ket{\Phi}}\mathscr{I}_{\ket{\Phi}}(\bm{\rho})\right]\equiv C_{\ket{\Phi}}(\bm{\rho})\,,
\end{equation}
where the optimization runs over all product states of $\mathcal{H}^{\otimes 2}$. Since any product state can be constructed from LU transformations on a reference product state $\ket{\Phi_0}\in \mathcal{H}^{\otimes 2}$, we can set
\begin{equation}
\ket{\Phi}=\left(\bigotimes_{n=1}^N \bm{V}_n \otimes \bigotimes_{n=1}^N \bm{W}_n\right)\ket{\Phi_0}\,,
\end{equation}
where $\bm{V}_n$ and $\bm{W}_n$ are $SU(d_n)$ elements. In this case, the optimization is performed over $2N$ independent (special) unitary transformations ($\bm{V}_n$ and $\bm{W}_n$), in terms of which the objective function takes the form
\begin{equation}
\mathscr{I}_{\ket{\Phi}}(\bm{\rho})=|\bra{\bm{0}}\overline{\bm{U}}_0^\dagger\,\bm{\rho}\, \bm{U}_0\ket{\bm{0}}|-\sum_{i=1}^{\mathfrak{n}}\sqrt{\bra{\bm{0}}{\bm{U}}_i^\dagger\,\bm{\rho}\, \bm{U}_i\ket{\bm{0}} \bra{\bm{0}}\overline{\bm{U}}_i^\dagger\,\bm{\rho}\, \overline{\bm{U}}_i\ket{\bm{0}}}\,,
\end{equation}
where we have chosen $\ket{\Phi_0}=\ket{0}^{\otimes N}\otimes \ket{0}^{\otimes N}\equiv\ket{\bm{0}}\otimes \ket{\bm{0}}$ and defined the LU transformations
\begin{equation}
\bm{U}_0\coloneq \bigotimes_{n=1}^N \bm{V}_n\,,\quad\overline{\bm{U}}_0\coloneq \bigotimes_{n=1}^N \bm{W}_n\,,\quad \bm{U}_i\coloneq \bigotimes_{n=1}^N\bm{S}_{i,n} \,,\quad\overline{\bm{U}}_i\coloneq \bigotimes_{n=1}^N \overline{\bm{S}}_{i,n}\,.
\end{equation}
In the above, the single particle unitary matrices $\bm{S}_{i,n}$ and $\overline{\bm{S}}_{i,n}$ have the index $i$ referring to a particular bipartition $\{A_i|B_i\}$ of $\mathcal{H}$ and, in terms of this, are given by:
\begin{equation}
\bm{S}_{i,n}\coloneq \left\{\begin{array}{cc}
\bm{W}_n&\mbox{if }n\in A_i\\
\bm{V}_n&\mbox{otherwise}
\end{array}
\right.\quad\mbox{and}\quad \overline{\bm{S}}_{i,n}\coloneq \left\{\begin{array}{cc}
\bm{V}_n&\mbox{if }n\in A_i\\
\bm{W}_n&\mbox{otherwise}
\end{array}
\right.\,.
\end{equation}
Thus, we end up with the optimization problem
\begin{equation}\label{eq:optproblemexact}
\max_{\bm{V}_n,\bm{W}_n \in SU(d_n)} |\bra{\bm{0}}\overline{\bm{U}}_0^\dagger\,\bm{\rho}\, \bm{U}_0\ket{\bm{0}}|-\sum_{i=1}^{\mathfrak{n}}\sqrt{\bra{\bm{0}}{\bm{U}}_i^\dagger\,\bm{\rho}\, \bm{U}_i\ket{\bm{0}} \bra{\bm{0}}\overline{\bm{U}}_i^\dagger\,\bm{\rho}\, \overline{\bm{U}}_i\ket{\bm{0}}}\,.
\end{equation}

Let us now restrict the analysis of this optimization problem to the case of $N$-qubits (i.e., $d_n=2$ for every $n=1,\ldots, N$). Since single-qubit unitary transformation require only $2$ real parameters and our objective function involves $2N$ unitaries, the number of real variables to be optimized is $4 N$. Although the linear scaling with $N$ is certainly an appealing feature of this approach, the resulting objective function is not everywhere differentiable. This fact compromises the performance of numerical algorithms that rely on differentiating the objective function to approach the optima.

\section{X-Heuristic}\label{sec:heuristic}
We now introduce a heuristic approach for estimating the GM-concurrence of a density matrix of $N$-qubits. To motivate our heuristic, we start by reviewing the concept of X-density matrices and how to quantify their GM-concurrence. An $N$-qubit $X$-density matrix is any density matrix that, in the computational basis, takes the form
\begin{equation}\label{eq:NqubitXstate}
\bm{\rho}_X=\left[\begin{array}{cccccccc}
a_1    &       &       &       &  &       &   &r_1e^{i\phi_1}\\
       &a_2    &       &       &  &       &r_2e^{i\phi_2} &   \\
       &       &\ddots &       &  &\iddots&   &   \\
       &       &       &a_n    &r_ne^{i\phi_n}&       &   &   \\
       &       &       &r_ne^{-i\phi_n}&b_n&       &   &   \\
       &       &\iddots&       &  &\ddots &   &   \\
       &r_2e^{-i\phi_2}&        &       &  &       &b_2&   \\
r_1e^{-i\phi_1}&       &        &       &  &       &   &b_1
\end{array}\right]
\end{equation}
with $n=2^{N-1}$ and, for every $k=1,\ldots,n$, $\{a_k, b_k\} \in \mathbb{R}_+$, $\phi_k\in[0,2\pi]$, $r_k \in [0,\sqrt{a_k b_k}]$ (non-negativity of $\bm{\rho}_X$) and $\sum_{k} (a_k+b_k)=1$ (normalization of $\bm{\rho}_X$). As shown by Hashemi Rafsanjani \emph{et al.}~\cite{12Rafsanjani062303},  such states can have their GM-concurrence easily  --- and exactly --- quantified by the formula:
\begin{equation}\label{eq:cx}
C_X(\bm{\rho}_X)=\max\left[0,c_X(\bm{\rho}_X)\right]
\end{equation}
where
\begin{equation}\label{eq:cXSn}
 c_X(\bm{\rho}_X)\coloneq 2\max_{k\in \mathfrak{S}_n}\left[r_k-\sum_{j\in \mathfrak{S}_n\setminus\{k\}}\sqrt{a_jb_j}\right]\quad\mbox{and}\quad \mathfrak{S}_n\coloneq\{1,\ldots,n\}\,.
\end{equation}
The proposed heuristic takes advantage of the simplicity of Eq.~(\ref{eq:cx}) to produce computable estimates of GM-concurrence for non-X density matrices. In fact, as noted in Ref.~\cite{12Rafsanjani062303} (and explictly demonstrated in the~\ref{app:xlbGM}), the evaluation of $C_X(\bm{\rho})$ for an arbitrary $N$-qubit density matrix $\bm{\rho}$ (written in the computational basis), leads to a lower bound on the GM-concurrence of $\bm{\rho}$. In principle, depending on the particular form of $\bm{\rho}$, this lower bound can be very loose and, consequently, a very poor estimate of $C_{GME}(\bm{\rho})$ [see, e.g., Eqs.~(\ref{eq:cxrhos3}) and~(\ref{eq:cxrhos4})]. However, since this lower bound is \emph{saturated} for X-density matrices, one should expect that the ``closest'' $\bm{\rho}$ is to a X-density matrix, the better that estimate will turn out to be.

The X-heuristic materializes this reasoning by ``approximating'' $\bm{\rho}$ to a X-matrix with a LU transformation $\bm{U}_l$ and, subsequently, estimating its GM-concurrence as $C_X(\tilde{\bm{\rho}})$, where
\begin{equation}\label{eq:rhotil}
\tilde{\bm{\rho}}\coloneq\bm{U}_l\,\bm{\rho}\,\bm{U}_l^\dagger\,.
\end{equation}
Of course, the constraint of LU transformation is sufficient to imply that $\bm{\rho}$ and $\tilde{\bm{\rho}}$ have the same amount of GM-concurrence. However, whereas $C_{GME}(\bm{\rho})$ is invariant under LU transformations of $\bm{\rho}$, $C_X(\bm{\rho})$ is not.

Once again, we are left with the problem of choosing a LU transformation, but now aiming to implement the ``approximation to the X-form''. In order to formalize the meaning of this \emph{approximation}, we introduce the penalty function
\begin{equation}\label{eq:f}
f(\tilde{\bm{\rho}})=\sum_{\substack{j>i\\ j+i\neq 2^N +1}}|\tilde{\rho}_{i,j}|^2\,,
\end{equation}
which (i) vanishes if and only if $\tilde{\bm{\rho}}$ is a X-density matrix and (ii) smoothly increases as the magnitudes of the off-X entries of $\tilde{\bm{\rho}}$ increase. Under this ``$f$-metric''\footnote{Here and throughout, the term ``metric'' is used in a loose way to suggest a notion of distance between a given density matrix and the set of X-density matrices.}, the optimal choice of $\bm{U}_l$ is given by the solution of the optimization problem
\begin{equation}\label{eq:optprobx}
\min_{\bm{U}_l} f(\bm{U}_l\, \bm{\rho}\, \bm{U}_l^\dagger)\,,
\end{equation}
which is a \emph{non-linear least square problem} on $2N$ real variables with residuals given by $|\tilde{\rho}_{i,j}|$. We can think of it as finding the LU related density matrix that most accurately ``fits'' the X-density matrix model. 

Besides halving the number of variables in the optimization problem~(\ref{eq:optproblemexact}), the objective function in (\ref{eq:optprobx}) is smooth, which implies that numerical algorithms can take advantage of well-defined derivatives to more efficiently approach a minimum. On the analytical side, for density matrices $\bm{\rho}$ with enough symmetry, the objective function $f(\bm{U}_l\bm{\rho}\bm{U}_l^\dagger)$ will be sufficiently simple to enable educated guesses on the location of the minima, which can be certified \emph{a posteriori} by evaluating the gradient and the spectrum of the Hessian matrix at the candidate point. In any case, even if the gradient vanishes and the Hessian turns out to be positive definite, we cannot be sure that we have attained a \emph{global} minimum. Nevertheless, at this point we may have already obtained a non-trivial lower bound on the amount of GM-concurrence of $\bm{\rho}$ at a considerably low computational cost.

\section{Applications}\label{sec:applications}
In this section we apply the X-heuristic to estimate the amount of GM-concurrence in two families of $N$-qubit density matrices: the diagonal symmetric states and the steady states of the driven Dicke Model at zero temperature. For each family, we benchmark the quality of the resulting estimates against those produced by the optimal-$\ket{\Phi}$ scheme.

\subsection{Diagonal symmetric states}

Diagonal symmetric states are a natural extension of W-states to the domain of multiqubit mixed states and have recently attracted much attention in the fields of GM-entanglement detection and quantification~\cite{09Toth170503,12Tura060302R,13Novo012305,14Quesada052319,14Wolfe140402}. Formally, they can be defined as
\begin{equation}\label{eq:defdss}
\bm{\rho}_{ds,N}\coloneq\sum_{k=0}^N p_k \ket{D_k^N}\!\bra{D_k^N}\,,
\end{equation}
where $p_k\in[0,1]$ and $\sum_k{p_k}=1$. Here, the states $\ket{D_k^N}$ are the (totally symmetric) $N$-qubit Dicke states of $k$ excitations~\cite{54Dicke99,03Stockton022112}, defined in the computational basis as
\begin{equation}
\ket{D_k^N}\coloneq\frac{1}{\sqrt{C_k^N}}\sum_{\sigma}\ket{\sigma(1,\stackrel{k}{\cdots},1,0,\stackrel{N-k}{\cdots},0)}
\end{equation}
with the summation running over every distinct permutation of the sequence of $k$ ones and $N-k$ zeros. Of course, there are $C_k^N=N!/[k! (N-k)!]$ such permutations, justifying the normalization constant upfront. Moreover, note that $\ket{D_1^N}$ are the $N$-qubit W-states~\cite{00Dur062314}, which sets the context in which $\bm{\rho}_{ds,N}$ can be considered a generalization thereof. In the~\ref{app:matrixform}, we give explicit matrix forms of $\bm{\rho}_{ds,N}$ in the computational basis for $N=2,3,4$.

\begin{figure}[h]
\centering
\includegraphics{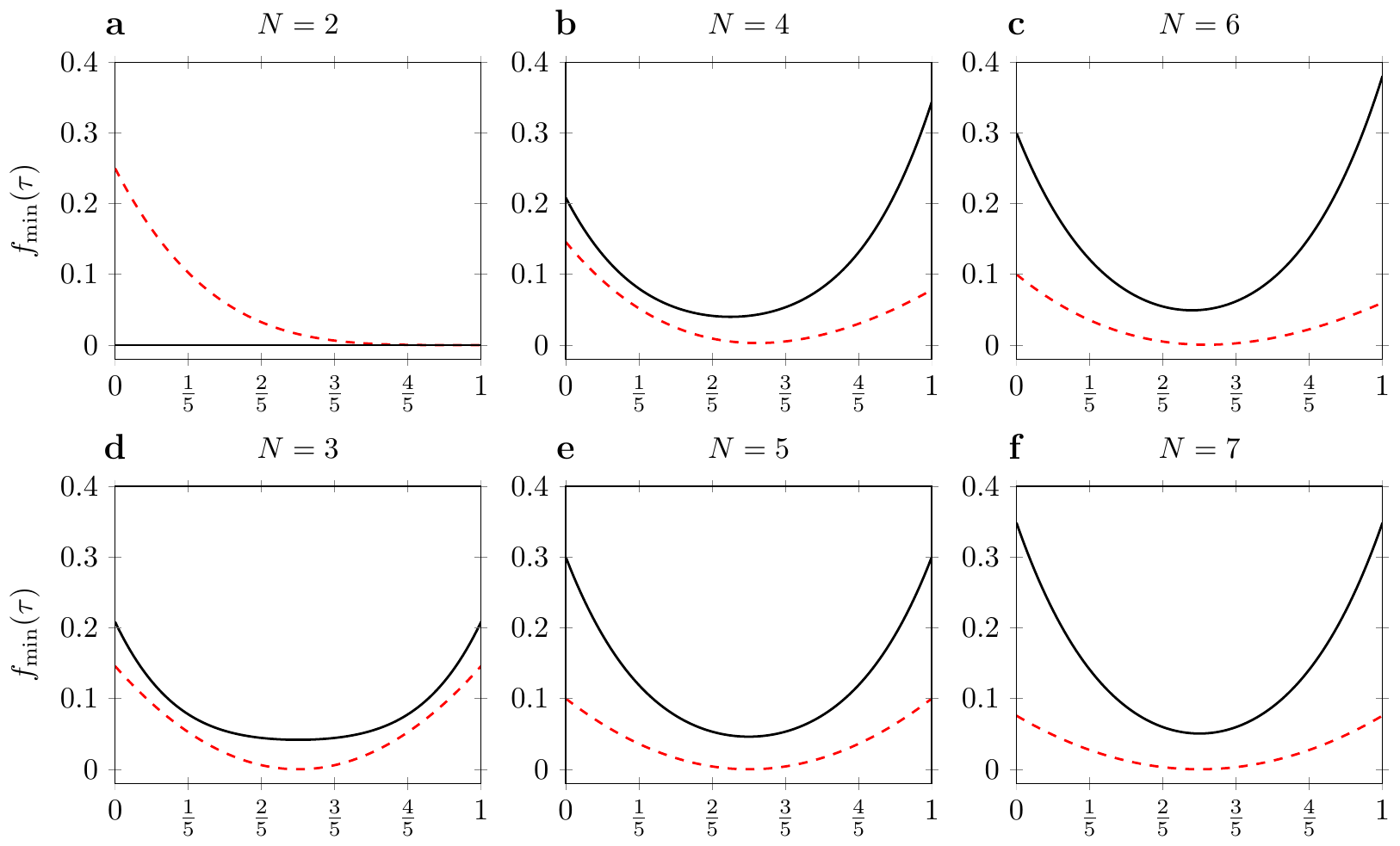}
\caption{Distance (``$f$-metric'') between the set of X-density matrices and the LU transformed diagonal symmetric state as a function of $\tau$.  The included dashed (red) lines represent (half of) the sum of the absolute values squared of the anti-diagonal terms. For $N=2$, the density matrix is of the X-form for every value of $\tau\in[0,1]$ and, as $\tau$ increases, it approximates a diagonal matrix. For $N>2$ the density matrices are not of the X-form for any value of $\tau\in[0,1]$ and better approximate this condition at intermediate values of $\tau$. 
}\label{fig:SYMfmin}
\end{figure}

For $N=2$, the resulting density matrix is already an X-state [cf. Eq.~(\ref{eq:rhodss2})]. As a result, $\bm{U}_l=\openone_4$, $f_{min}=0$ and $C_X(\bm{\rho}_{ds,2})=C_{GME}(\bm{\rho}_{ds,2})$. For $N>2$, the corresponding density matrices are no longer of the X-form [see, e.g., Eqs.~(\ref{eq:rhodss3}) and~(\ref{eq:rhodss4})] and thus, in order to optimally estimate $C_{GME}$ with $C_{X}$, we must first determine a LU transformation 
\begin{equation}
\bm{U}_l=\bigotimes_{j=1}^N\left[\begin{array}{cc}
\cos\vartheta_j & \sin \vartheta_j e^{i\varphi_j}\\
-\sin \vartheta_j e^{-i\varphi_j} & \cos \vartheta_j
\end{array}
\right]
\end{equation}
that minimizes $f(\bm{U}_l\bm{\rho}_{ds,N}\bm{U}_l^\dagger)$. Thanks to the symmetry of $\bm{\rho}_{ds,N}$, some simple analysis of this function reveals a minimum at\footnote{A complementary numerical analysis strongly suggests that Eq.~(\ref{eq:SYMXminimumcoords}) gives, actually, a global minimum.}
\begin{equation}\label{eq:SYMXminimumcoords}
\vartheta_j=\frac{\pi}{4}\quad\mbox{and}\quad \varphi_j=0\quad \mbox{for every}\quad  j \in \mathfrak{S}_N\,,
\end{equation}
for which the corresponding values of $f=f_{min}$ can be promptly computed with the aid of Eqs. (\ref{eq:rhotil}) and (\ref{eq:f}). In Fig.~\ref{fig:SYMfmin}, these are plotted for a family of populations $p_k$ specified by the single parameter $\tau\in[0,1]$ as follows:

\begin{equation}\label{eq:dssparams}
 p_{\left\lfloor\frac{N}{2}\right\rfloor}=(\tau-1)^2\,,\quad p_{\left\lfloor\frac{N}{2}\right\rfloor+1}=\tau^2\quad\mbox{and}\quad p_\ell= \frac{2\tau(1-\tau)}{N-1} \quad \mbox{for every }\ell\in \{0,\ldots,N\}\setminus\left\{\left\lfloor\frac{N}{2}\right\rfloor, \left\lfloor\frac{N}{2}\right\rfloor+1\right\}\,.
 \end{equation}

Intuitively, the plots of Fig.~\ref{fig:SYMfmin} can be interpreted as some error measure associated with the process of estimating $C_{GME}(\bm{\rho}_{ds,N})$ as $C_X(\tilde{\bm{\rho}}_{ds,N})$. Indeed, if $f_{min}=0$ (e.g., case $N=2$), then $C_X(\tilde{\bm{\rho}}_{ds,N})$ is exactly equal to $C_{GME}(\bm{\rho}_{ds,N})$. As $f$ increases, $\tilde{\bm{\rho}}_{ds,N}$ deviates from the X-form and, hence, the estimate $C_X(\tilde{\bm{\rho}}_{ds,N})$ is expected to be poorer. Nevertheless, recall that $C_X(\tilde{\bm{\rho}}_{ds,N})$ is always a lower bound on $C_{GME}(\bm{\rho}_{ds,N})$, so the ``$f$-error bars'' are not \emph{centered} in $C_X(\tilde{\bm{\rho}}_{ds,N})$, but lay strictly \emph{above} it.

\begin{figure}[h]
\centering
\includegraphics{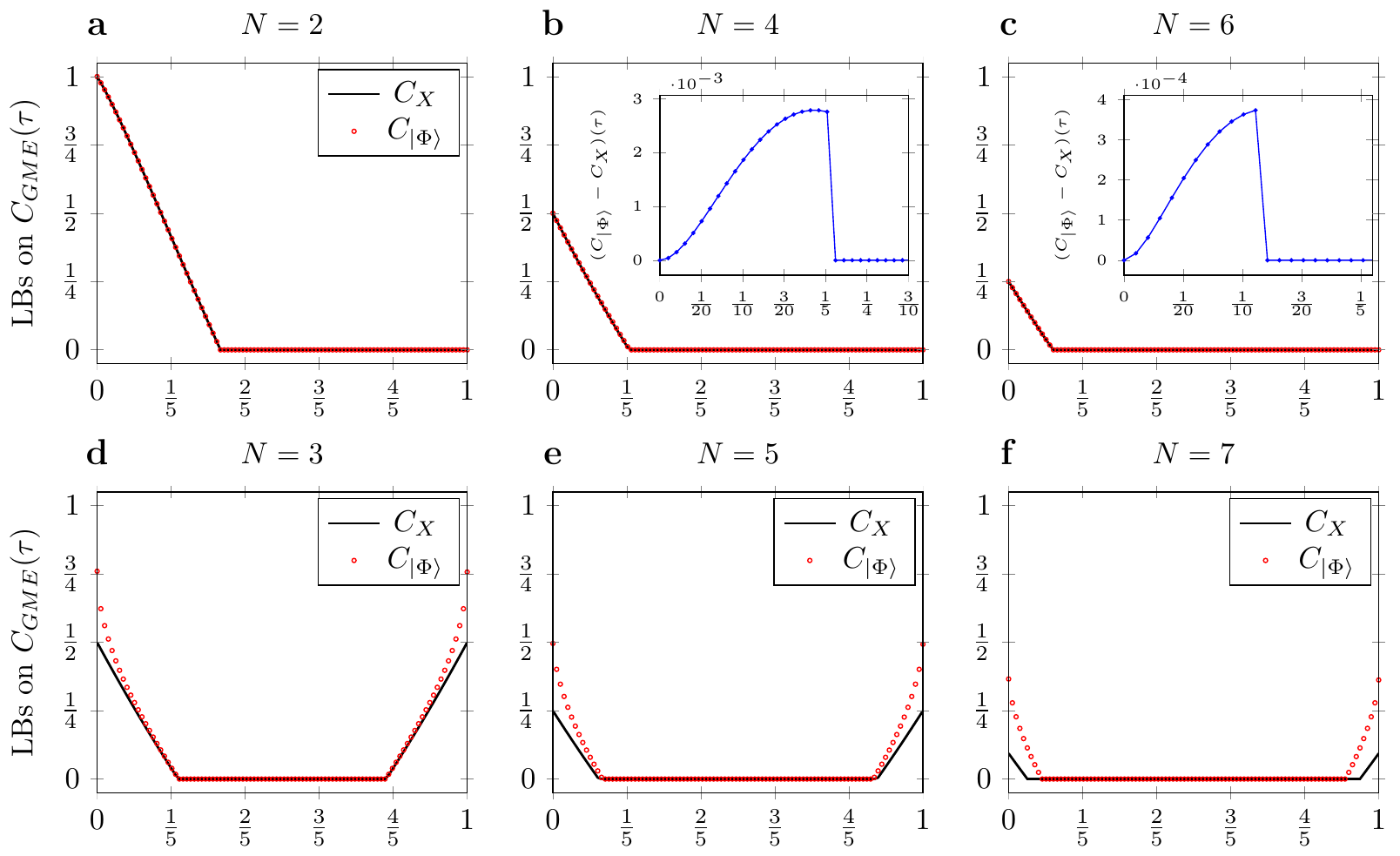}
\caption{Comparison between lower bounds (LBs) on the GM-concurrence of a family of $N$-qubit diagonal symmetric states [cf. Eq.~(\ref{eq:dssparams})]. The solid line represents $C_X(\tilde{\bm{\rho}}_{ds,N})$ and the red circles represent $C_{\ket{\Phi}}(\bm{\rho}_{ds,N})$. While the density matrices $\tilde{\bm{\rho}}_{ds,N}$ (and their corresponding X-concurrence) were analitically computed [cf. Eqs.~(\ref{eq:defdss})--(\ref{eq:dssparams})], the values of $C_{\ket{\Phi}}(\bm{\rho}_{ds,N})$ were obtained from a numerical implementation of the optimal $\ket{\Phi}$ method [cf. Eq.~(\ref{eq:optproblemexact})].}\label{fig:SYMlowerbound}
\end{figure}

In Fig.~\ref{fig:SYMlowerbound} we present the resulting values of $C_X(\tilde{\bm{\rho}}_{ds,N})$ (solid line) along with the points corresponding to $C_{\ket{\Phi}}(\bm{\rho}_{ds,N})$ for one hundred values of $\tau$ uniformly distributed in the range $[0,1]$. While the line was analytically constructed, each point was obtained by numerically solving the optimization problem~(\ref{eq:optproblemexact}). Overall, we notice that both estimates follow very similar trends and coincide in a significant portion of the domain. Next, their similarities and differences are discussed in greater detail.

For $N=2$, owing to the X-form of $\bm{\rho}_{ds,2}$, $C_X$ is precisely equal to the Wootters concurrence and, therefore, exactly matches $C_{\ket{\Phi}}$ for every $\tau\in[0,1]$ --- see Plot~\ref{fig:SYMlowerbound}a. For $N=4, 6$, this \emph{exact} matching no longer occurs for $0<\tau\lesssim 0.212$ ($N=4$) and $0<\tau \lesssim 0.121$ ($N=6$). Nevertheless, as the insets in Plots~\ref{fig:SYMlowerbound}b,c show, the difference between $C_{\ket{\Phi}}$ and $C_X$ is orders of magnitude smaller than the actual value of the estimates, being thus negligible for most practical applications. For odd $N$, a more significant mismatch between $C_X$ and $C_{\ket{\Phi}}$ occurs at the extremes of the domain (cf. Plots~\ref{fig:SYMlowerbound}d,e,f), but it rapidly decreases and vanishes in a large central region of the domain, where neither estimates detect GM-entanglement. An important difference between $C_X(\tilde{\bm{\rho}}_{ds,N})$ and $C_{\ket{\Phi}}(\bm{\rho}_{ds,N})$ becomes apparent in the case $N=7$ (Plot~\ref{fig:SYMlowerbound}f), where we note that each estimate arrives to (and departs from) zero at slightly different values of $\tau$, in such a way that there are a few states whose GM-entanglement can be detected by the optimal-$\ket{\Phi}$ scheme, but not by the X-heuristic.

\subsection{Steady states of the driven Dicke model}
From the theoretical point of view, the Dicke model~\cite{54Dicke99} represents a special quantum mechanical model whose unique mathematical and physical properties allow us to describe important cooperative phenomena (such as, for
example, resonance fluorescence and cooperative emission) far from the thermodynamic limit. Their virtues are also reflected in works involving the
exact steady state solutions of its master equation that include (or not) the effects of detuning between a collective driving field and the atomic
resonant frequency~\cite{79Puri200,80Drummond1179,81Lawande4171,02Schneider042107}. In this section, we focus on the steady state solution of the driven Dicke model at zero temperature

\begin{equation}\label{eq:rhosN}
\bm{\rho}_{s,N}=\frac{1}{D_N}\sum_{m,n=0}^N\left(\frac{\bm{J}_-}{g^\ast}\right)^m\left(\frac{\bm{J}_+}{g}\right)^n\,,
\end{equation}
where $N$ depicts the number of two-level atoms (ions), $D_{N}$ is a normalization constant, $\gamma \coloneq \gamma_{A}/\Omega$ denotes the ratio between the Einstein $A$-coefficient $\gamma_{A}$ of each atom (ion) and the Rabi frequency $\Omega$, and $g \coloneq i/\gamma$. Note that $\bm{J}_{\pm} \coloneq \sum_{\ell} \openone_{2^{\ell-1}}\otimes\bm{\sigma}_\pm\otimes\openone_{2^{N-\ell}}$ for $\ell=1,\ldots,N$ correspond to the collective raising and lowering operators expressed in terms of the Pauli matrices $\bm{\sigma}_{\pm}$ which satisfy, together
with the collective inversion operator $\bm{J}_{z}$, the usual angular momentum commutation relations.

\begin{figure}[h]
\centering
\includegraphics{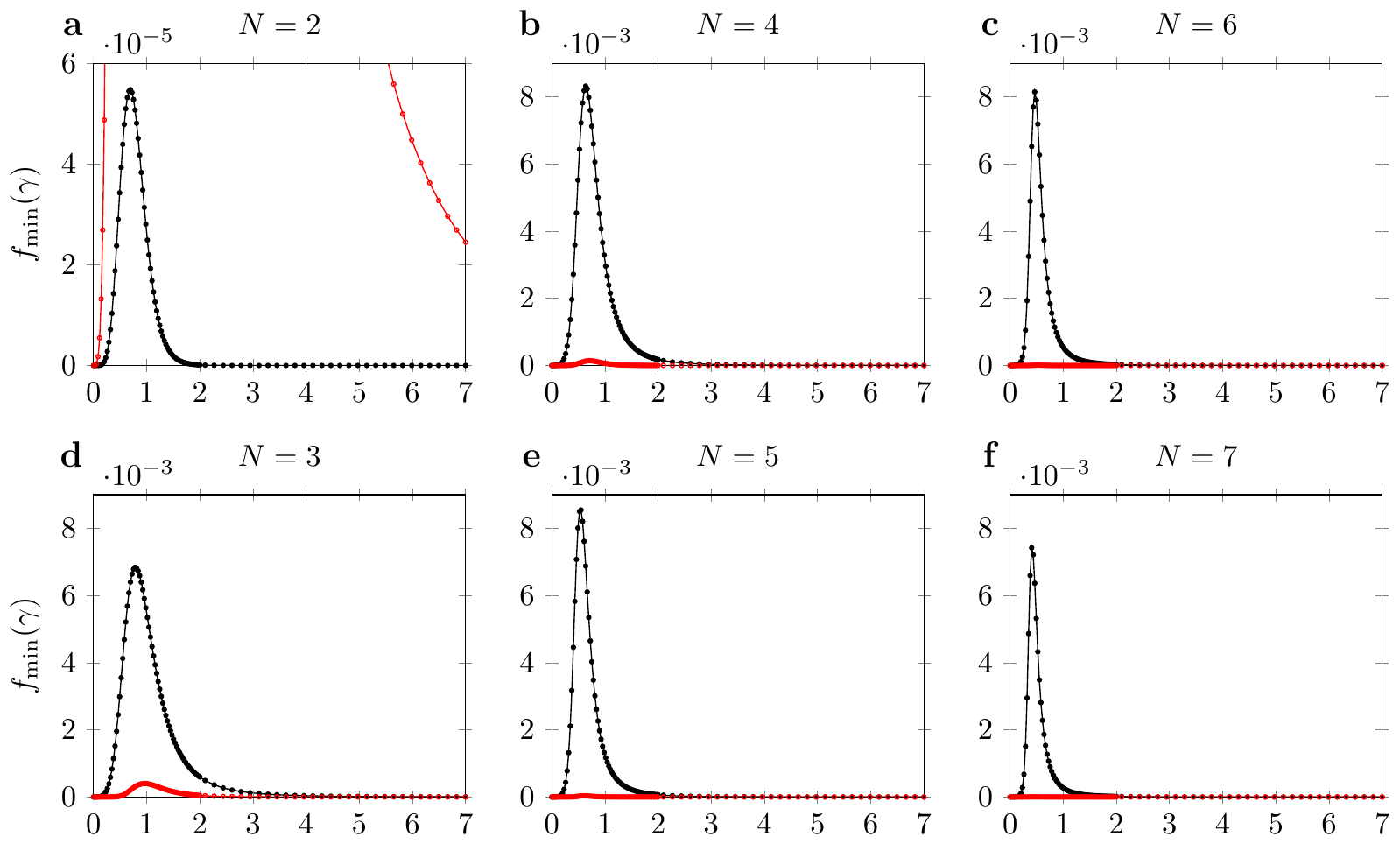}
\caption{Distance (``$f$-metric''), as function of $\gamma$, between the set of X-density matrices and the closest LU transformed steady state of the driven Dicke model. The included red data points represent (half of) the sum of the absolute values squared of the anti-diagonal terms. For $N>2$, the LU operation that optimally decreases the off-X entries has a deleterious effect on the anti-diagonal as well, essentially making diagonal the transformed density matrix.}\label{fig:minfvsgamma}
\end{figure}

\begin{figure}[h]
\centering
\includegraphics{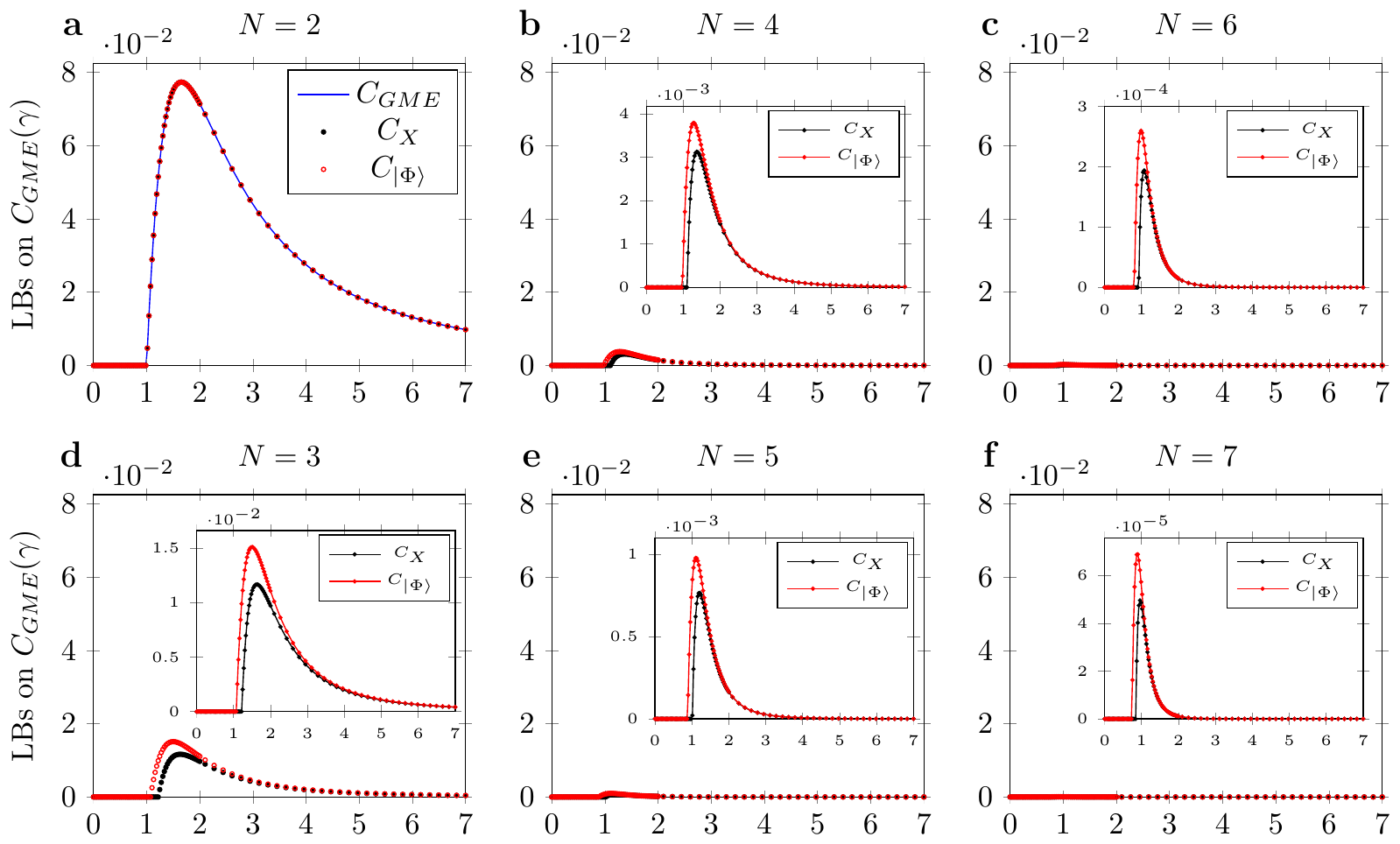}
\caption{Comparison between the lower bounds $C_X(\tilde{\bm{\rho}}_{s,N})$ (black circles) and $C_{\ket{\Phi}}(\bm{\rho}_{s,N})$ (red circles) of the genuine multipartite concurrence of the $N$-qubit steady states of the driven Dicke model [cf. Eq.~(\ref{eq:rhosN})]. Both estimates were numerically calculated.}\label{fig:DIKlowerbound}
\end{figure}

In the~\ref{app:matrixform}, explicit matrix forms of $\bm{\rho}_{s,N}$ in the computational basis are given for $N=2,3,4$. Noticeably, the resulting matrices are not of the X-form and, as shown in Fig.~\ref{fig:minfvsgamma}, they cannot be exactly brought to the X-form via LU transformations (i.e., they do not assume the X-form in any orthonormal basis of product states). These observations qualify $\bm{\rho}_{s,N}$ as a good testbed for the X-heuristic and so, in Fig.~\ref{fig:DIKlowerbound}, we present the corresponding estimates obtained from numerical implementations of the X-heuristic and optimal-$\ket{\Phi}$ scheme.

At a glance of Fig.~\ref{fig:DIKlowerbound}, both $C_X$ and $C_{\ket{\Phi}}$ are seen to follow very similar trends: a zero plateau that extends from $\gamma=0$ up until $\gamma\approx 1$, followed by a sudden growth and, finally, an asymptotic decay. In what follows we offer a more in-depth analysis of these results by considering the cases $N=2$ and $N>2$ separately.

In the case $N=2$, on top of $C_X$ and $C_{\ket{\Phi}}$, we have also plotted the Wootters concurrence of $\bm{\rho}_{s,2}$ which, in this case, is the ``genuine multipartite'' concurrence $C_{GME}$. Despite $\tilde{\bm{\rho}}_{s,2}$ not being of the X-form, we note that the three measures coincide up to the numerical precision of $10^{-14}$, with the zero plateu ranging up until $\gamma=1.00$ and the maximal concurrence ($\sim 7.735\times 10^{-2}$) occurring at $\gamma_{max}\approx 1.65$. Interestingly, the observed coincidence between the three measures signals that the X-concurrence formula may also be an exact concurrence formula for some non-X states. In~\ref{app:xlbGM}, this point is further explored with a demonstration that $C_X(\bm{\rho}_{s,2})$ matches \emph{exactly} the Wootters concurrence ($C_W$) of $\bm{\rho}_{s,2}$. This observation provides a (constructive) negative answer to an open question in Ref.~\cite{12Rafsanjani12043912} on whether the saturation of $C_{W}(\bm{\rho})\geq C_X(\bm{\rho})$ requires $\bm{\rho}$ to be of the X-form in some orthonormal product basis.

For $N>2$, $C_X$ and $C_{\ket{\Phi}}$ no longer coincide for every value of $\gamma$. Nevertheless, the difference between them asymptoticaly decreases and nearly vanishes at relatively small values of $\gamma$. For example, the ratio $(C_{\ket{\Phi}}-C_X)/C_{\ket{\Phi}}$ is smaller than $3.2\%$ for $\gamma\gtrsim 7$ ($N=3$), $\gamma\gtrsim 2.2$ ($N=4$), $\gamma\gtrsim 1.66$ ($N=5$), $\gamma\gtrsim 1.4$ ($N=6$) and $\gamma\gtrsim 1.2$ ($N=7$). Of course, this behavior was to be expected since for such values of $\gamma$ we have already seen, in Fig.~\ref{fig:minfvsgamma}, that $\tilde{\bm{\rho}}_{s,N}$ is nearly X-formed. In fact, thanks to that, we can also infer that $C_X$ must well aproximate the (unknown) value of $C_{GME}$ in such values of $\gamma$.

As it ocurred with the diagonal symmetric states, the main drawback in replacing $C_{\ket{\Phi}}$ with $C_X$ is that there are certain states that have its GM-entangled detected by the former but not by the latter. In Fig.~\ref{fig:DIKlowerbound}, this is expressed  by the fact that the zero plateau generated by the X-heuristic is longer than that obtained with the optimal-$\ket{\Phi}$ scheme. However, as the plots show, the difference is usually very small, meaning that failure to detect GM-entanglement with the X-heuristic will only occur for very few and specific values of $\gamma$. Other than that, it is also worth noticing that each method foresees a maximal value of $C_{GME}$ at slightly different values of $\gamma$, in such a way that $\gamma_{max}^{(X)}>\gamma_{max}^{(\ket{\Phi})}$. However, the difference $\gamma_{max}^{(X)}-\gamma_{max}^{(\ket{\Phi})}$ is again very small and, ultimately, it is not clear which one more accurately describes the actual location of the maximum of $C_{GME}$.

\section{Computational Efficiency}\label{sec:efficiency}
The GM-concurrence estimates produced by the X-heuristic cannot improve on the corresponding estimates produced by the optimal-$\ket{\Phi}$ scheme; a fact that follows directly from the derivation of the X-heuristic (cf. Sec.~\ref{sec:heuristic}). There is, however, an important trade-off between accuracy and efficiency that must be considered before disregarding the X-heuristic in favor of the optimal-$\ket{\Phi}$ scheme. On the accuracy side, we have already seen that the X-estimates are usually pretty good approximations of the optimal-$\ket{\Phi}$ estimates. On the efficiency side, we now explicitly demonstrate that the X-estimates are significantly easier to compute both for the diagonal symmetric states and for the steady states of the driven Dicke model.

As mentioned before, the computational advantage of the X-heuristic over the optimal-$\ket{\Phi}$ scheme relies upon two aspects of the objective function associated with the former: (i) it is smooth and (ii) it depends on a number of variables that halves the number of variables in the (non-smooth) objective function of the optimal-$\ket{\Phi}$ scheme. While the relevance of (ii) is obvious, the importance of (i) cannot be overestimated. Thanks to (i), the X-heuristic optimization problem can be regarded as ideal for the application of a \emph{quasi-Newton} method~\cite{06Nocedal}, a family of nonlinear optimization algorithms that take advantage of well-defined derivatives of smooth objective functions to more efficiently converge to a minimum. 

In this work, the numerical computations of the X-heuristic and optimal-$\ket{\Phi}$ estimates were performed with the MATLAB function \texttt{fminunc}, which implements a popular quasi-Newton algorithm known as BFGS (after its discoverers Broyden, Fletcher, Goldfarb and Shanno). As with any quasi-Newton algorithm, the BFGS performs successive evaluations of the gradient to build a quadratic model of the objective function that is sufficiently good to attain a superlinear rate of convergence. It contrasts with Newton's method~\cite{06Nocedal} in the sense that it does not require (nor attempts to compute) the Hessian matrix, typically a time-consuming and error-prone task. Instead, at each step, it gains information about the second derivative along the search direction by considering changes in the gradient. Although the BFGS algorithm requires repeated computations of the gradient, it has been noted to perform well also in nonsmooth optimization problems, as long as it does not run into a nonsmooth point (see, e.g.,~\cite{13Lewis135} and references therein). Consistently with this observation, we have observed a better performance of the optimal-$\ket{\Phi}$ scheme with the BFGS algorithm than with algorithms that do not rely on evaluations of the gradient (e.g., Nelder Mead algorithm~\cite{65Nelder308}, implemented by the MATLAB function \texttt{fminsearch}), especially for larger values of $N$. For this reason, aiming to build the most efficient implementation of the two schemes, we employed the BFGS algorithm for both. 

In order to compare the performance of the two schemes in producing reasonable estimates of GM-concurrence, we set a threshold equal to the X-estimates presented in Figs.~\ref{fig:SYMlowerbound} and \ref{fig:DIKlowerbound} and timed how long it took for each scheme to reach that threshold. As it is usually the case with search algorithms, a starting value had to be provided, and so we resorted to random initial guesses (uniformly sampled) to avoid biasing the optimizer toward (or against) a satisfactory minimum. By doing so, we could gauge how hard it is for each scheme to reach the threshold from a zero-knowledge initial condition. 

\begin{figure}[h]
\centering
\includegraphics{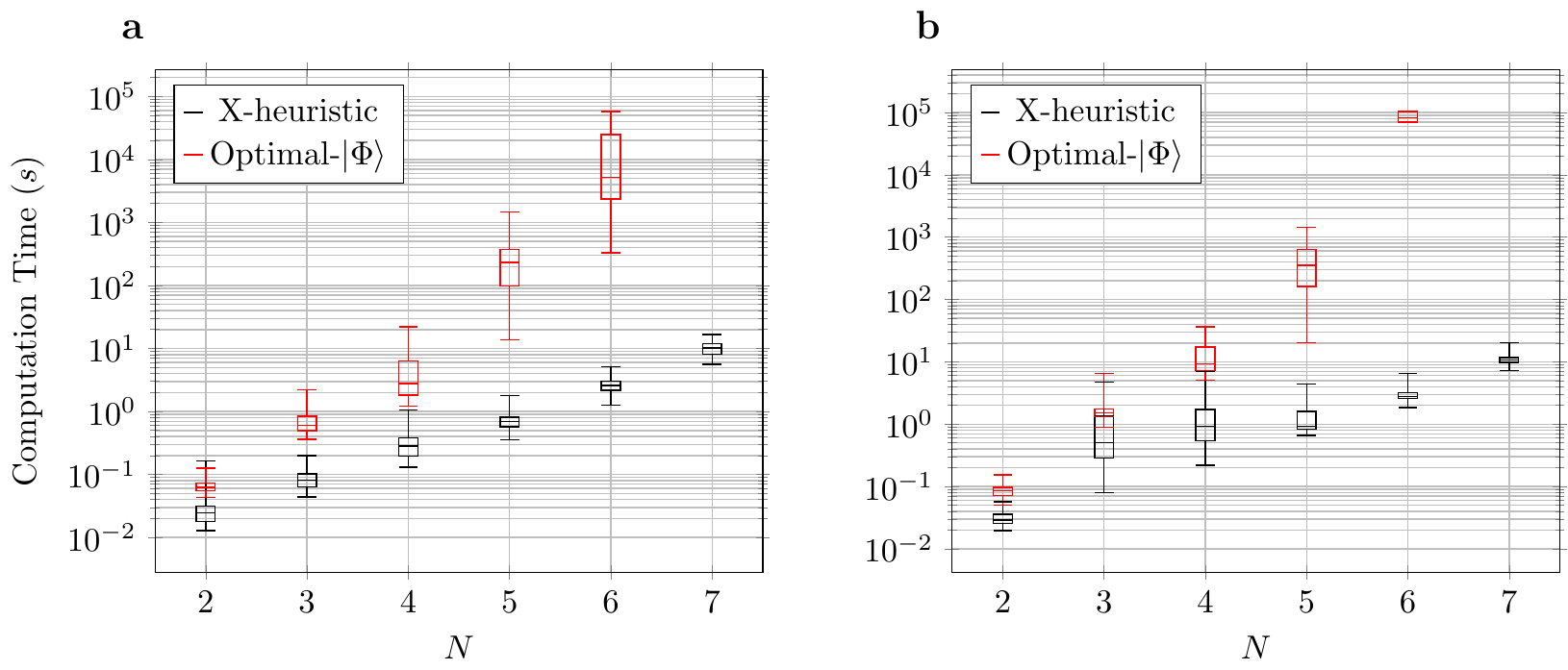}
\caption{Box-and-whisker plots of computation time taken to reach the X-estimates of GM-concurrence shown in Figs.~\ref{fig:SYMlowerbound} and \ref{fig:DIKlowerbound} for (a) the $N$-qubit diagonal symmetric states with $\tau=3.03\times 10^{-2}$ and (b) the $N$-qubit steady state of the driven Dicke model (zero temperature) corresponding to the peaks of the X-estimate in Fig.~\ref{fig:DIKlowerbound}, i.e., $\gamma=1.652$ ($N=2$), $\gamma=1.623$ ($N=3$), $\gamma=1.362$ ($N=4$), $\gamma=1.217$ ($N=5$), $\gamma=1.072$ ($N=6$), $\gamma=0.956$ ($N=7$). The time measurements were conducted while running our MATLAB implementations of the X-heuristic (black) and optimal-$\ket{\Phi}$ scheme (red) with random initial guesses (uniformly sampled) and termination tolerances on the parameter values (\texttt{TolX}) and on the objective function value (\texttt{TolFun}) set to $10^{-11}$. For each combination of method and state, the timing was repeated $100$ times (except for the optimal-$\ket{\Phi}$ method applied to states with $N=6$, in which cases we contented ourselves with only $5$ repetitions). The plots display the five-number summary of the resulting time distributions. Whenever the optimization terminated without converging to a value equal to or greater than the desired threshold, another try was made (with a different random initial guess) until the threshold was reached. Computations were performed on a $2.2$-GHz Intel Core i7-2670QM.}\label{fig:time}
\end{figure}

Each optimization was repeated $100$ times\footnote{Except for the time measurements of the optimal-$\ket{\Phi}$ scheme for states with $N=6$, which were repeated only $5$ times due to the long time-frame necessary to converge to a satisfactory value.}, and the five-number summary of the resulting ``wall-clock time'' distributions is shown in Fig.~\ref{fig:time} as a box-and-whiskers plot. Plot~\ref{fig:time}a refers to the time measurements obtained for the diagonal symmetric state with $\tau=3.03\times 10^{-2}$ for every $N\in[2,7]$, whereas Plot~\ref{fig:time}b refers to the time measurements for the steady states of the driven Dicke model that, for each $N$, have the value of $\gamma$ corresponding to the peak of the X-estimates in Fig.~\ref{fig:DIKlowerbound}. We note that to obtain each box-and-whisker appearing in Fig.~\ref{fig:time}, several initial guesses were typically made because either (i) the optimizer converged to a unsatisfactory minimum or (ii) the optimizer failed to converge after $10^4$ iterations. Of course, the resulting box-and-whiskers take into account every (if any) unsuccessful attempt as well as the successful one.

Remarkably, Fig.~\ref{fig:time} reveals a significant advantage of the X-heuristic over the optimal-$\ket{\Phi}$ scheme, especially for the larger values of $N$. For $N=7$, for example, the X-heuristic estimates were produced in roughly $10$~s, whereas the optimal-$\ket{\Phi}$ scheme was unable to reach the threshold once, even after a week (approximately $6\times 10^5$~s) of computation. Most importantly, the plots suggest that the computational time for each scheme scales differently with $N$, with the X-heuristic resembling an exponential growth and the optimal-$\ket{\Phi}$ some worse scaling. Naturally, this observation cannot be explained by the difference in the number of variables associated to each scheme (since they both scale linearly with $N$), and we assign it to the smoothness (or lack thereof) of the objective functions. Because of this scaling, the X-heuristic may be the only viable option for GM-concurrence estimation in generic $N$-qubit states with intermediate values of $N$.

The results of this section indicate that, despite the theoretical superiority of the optimal-$\ket{\Phi}$ scheme over the X-heuristic, the latter may actually overperform the former in practical situations where a time horizon has to be considered. For example, if we allow one second to obtain a GM-concurrence estimate for a state with $N\in[3,5]$, then Fig.~\ref{fig:time} implies that we are much more likely to obtain a better estimate with the X-heuristic than with the optimal-$\ket{\Phi}$ scheme, which may have converged to a lesser estimate (or even have not converged at all). Naturally, time constraints like this become virtually more important as $N$ increases.

\section{Concluding Remarks}\label{sec:conclusion}
In this paper we introduced a heuristic method for estimating the GM-concurrence of an $N$-qubit density matrix. It consists of evaluating the $N$-qubit X-state GM-concurrence formula for density matrices that are not of the X-form, but were previously brought ``as close as possible'' to it with a LU transformation. We have shown that the estimates thus produced are lower bounds on the GM-concurrence by demonstrating that they actually lower bound a standard lower bound on the GM-concurrence (attainable via a numerical procedure which we refer to as the ``optimal-$\ket{\Phi}$'' method). Most importantly, by examining two prominent families of mixed $N$-qubit states, we found that our estimates are usually very close to those produced by the optimal-$\ket{\Phi}$ scheme and are significantly easier to be numerically computed and, in certain cases of high symmetry, can be analytically obtained.

As for future directions, let us mention two possible extensions of the present work. First, it would be interesting to evaluate the performance of the X-heuristic in other well-established families of $N$-qubit mixed states such as reduced ground states of spin chain models. As a matter of fact, it is fair to say that the X-heuristic has already been successfully applied in this context, in Ref.~\cite{14Giampaolo93033}, where the genuine tripartite concurrence of the reduced ground state of three spins symmetrically distributed in the cluster-Ising model was exactly calculated thanks to the fact that such states can be turned into X-states via LU transformations. Nevertheless, the application of our heuristic in alike models for which the X-form is not exactly attainable is yet to be explored and is likely to provide an efficient tool for GM-entanglement estimation in condensed matter systems.

Second, the requirement of approximating the X-form via LU transformations is unnecessarily strong for the goal of preserving GM-concurrence. In fact, it was recently conjectured that a generic two-qubit density matrix can always be turned into a X-density matrix of same entanglement via a unitary transformation that is not necessarily local~\cite{13Hedemann}. In Ref.~\cite{14Mendonca79}, this conjecture was rigourously proved for three entanglement measures and, moreover, a semi-analitic prescription for constructing the corresponding ``entanglement-preserving-unitaries'' was delineated. Along the same lines, it would be interesting to characterize the most general family of ``GM-concurrence-preserving-unitaries'' and attempt to approximate the X-form via conjugation with elements of this less constrained family. In that case, better X-form approximations should be expected, presumably improving the quality of the resulting GM-concurrence estimates. Of course, this new optimization problem may be less suitable for numerical implementation, in which case the trade-off between efficiency and accuracy should be carefully reconsidered.

\section*{Acknowledgements}
P.E.M.F.M. acknowledges the financial support of the Brazilian Air Force and the program ``Ci\^encia sem Fronteiras'', Project No. 200024/2014-0.

\appendix
\section{X-concurrence is a lower bound for GM-concurrence}\label{app:xlbGM}

In this appendix we demonstrate that the X-concurrence formula~(\ref{eq:cx}) applied to a generic $N$-qubit density matrix $\bm{\rho}$ is equal to $\max_{\mu} 2 \mathscr{I}_{\ket{\Phi_\mu}}(\bm{\rho})$ for a certain family of states $\{\ket{\Phi_\mu}\}_{\mu=0}^{n-1}$ (recall that $n\coloneq 2^{N-1}$), being thus a lower bound on the GM-concurrence.

Consider the following (decimal) representation for the $N$-qubit computational basis
\begin{align}
\ket{0}\equiv\ket{0\,0\,\cdots\,0\,0}\,,\quad \ket{1}\equiv\ket{0\,0\,\cdots\,0\,1}\,,\,\ldots\,,\quad \ket{N}\equiv\ket{1\,0\,\cdots\,0\,0}\,,\nonumber\\
 \ket{N+1}\equiv\ket{1\,0\,\cdots\,0\,1}\,,\,\ldots\,,\quad \ket{2^N-1}\equiv\ket{1\,1\,\cdots\,1\,1}\,,
\end{align}
and, in terms of this, define
\begin{equation}
\ket{\Phi_\mu}\coloneq \ket{\mu}\otimes\ket{2^N-1-\mu}\quad\mbox{for every}\quad \mu\in\{0,\ldots,n-1\}\,.
\end{equation}
 Substitution of this into Eq.~(\ref{eq:Iketphirho}) gives, after some straightforward algebra,
\begin{equation}\label{eq:Iketphijrho}
\mathscr{I}_{\ket{\Phi_\mu}}(\bm{\rho})=|\bra{\mu}\bm{\rho}\ket{2^N-1-\mu}|-\sum_{\nu\in \{0,\ldots,n-1\}\setminus\{\mu\}}\sqrt{\bra{\nu}\bm{\rho}\ket{\nu}\bra{2^N-1-\nu}\bm{\rho}\ket{2^N-1-\nu}}.
\end{equation}
Now, let the density matrix of $\bm{\rho}$ be decomposed (in the computational basis) as 
\begin{equation}\label{eq:NqubitXstate}
\bm{\rho}=\left[\begin{array}{cccccccc}
a_1    &       &       &       &  &       &   &r_1e^{i\phi_1}\\
       &a_2    &       &       &  &       &r_2e^{i\phi_2} &   \\
       &       &\ddots &       &  &\iddots&   &   \\
       &       &       &a_n    &r_ne^{i\phi_n}&       &   &   \\
       &       &       &r_ne^{-i\phi_n}&b_n&       &   &   \\
       &       &\iddots&       &  &\ddots &   &   \\
       &r_2e^{-i\phi_2}&        &       &  &       &b_2&   \\
r_1e^{-i\phi_1}&       &        &       &  &       &   &b_1
\end{array}\right]+\bm{\mathcal{O}}
\end{equation}
where $\bm{\mathcal{O}}$ is some $2^N\times 2^N$ Hermitian matrix with zeros along the main- and anti-diagonals. In terms of this, Eq.~(\ref{eq:Iketphijrho}) clearly takes the form
\begin{equation}
\mathscr{I}_{\ket{\Phi_\mu}}(\bm{\rho})=r_{\mu+1}-\sum_{\nu\in \{0,\ldots,n-1\}\setminus\{\mu\}}\sqrt{a_{\nu+1}b_{\nu+1}}.
\end{equation}
Making $\nu+1=j$ and $\mu+1=k$ (in such a way that $j,k\in \mathfrak{S}_n$), it is clear that 
\begin{equation}
\mathscr{I}_{\ket{\Phi_{k-1}}}(\bm{\rho})=r_{k}-\sum_{j\in \mathfrak{S}_n\setminus\{k\}}\sqrt{a_{j}b_{j}}
\end{equation}
and, thus [cf. Eq.~(\ref{eq:cXSn})]
\begin{equation}
c_X(\bm{\rho})=2\max_{k\in \mathfrak{S}_n}\mathscr{I}_{\ket{\Phi_{k-1}}}(\bm{\rho})\,.
\end{equation}
This fact, combined with inequality (\ref{eq:strongest}) and identity~(\ref{eq:cx}), leads to the desired result:
\begin{equation}\label{eq:lbCgmeCx}
C_{GME}(\bm{\rho})\geq \max\left[0,2\max_{\ket{\Phi}}\mathscr{I}_{\ket{\Phi}}(\bm{\rho})\right] \geq \max\left[0,2\max_{k\in \mathfrak{S}_n} \mathscr{I}_{\ket{\Phi_{k-1}}}(\bm{\rho})\right]=\max[0,c_X(\bm{\rho})]=C_X(\bm{\rho})\,.
\end{equation}

As suggested in Ref.~\cite{12Rafsanjani12043912} (within the framework of $N=2$), it is interesting to look for density matrices that saturate the above inequalities. Of course, saturation occurs if $\bm{\rho}$ is an X-state, other than that (and to the best of our knowledge), no examples of density matrices saturating inequality~(\ref{eq:lbCgmeCx}) have been identified so far. In what follows, we introduce a family of two-qubit density matrices which are not of the X-form in any product state basis and, nonetheless, satisfy $C_{W}(\bm{\rho})=C_X(\bm{\rho})$, where $C_W$ denotes the Wootters concurrence (i.e., the ``genuine multipartite'' concurrence of two-qubit states).

Consider the two-qubit steady states of the driven Dicke model at zero temperature, whose density matrix in the computational basis is presented in Eq.~(\ref{eq:ss}). Noticeably, it is not an X-state for $\gamma\neq 0$ and, as Plot~\ref{fig:minfvsgamma}a shows, there is no LU transformation capable of simulatenously vanishing all of its non-X entries (although this can be remarkably well aproximated for sufficiently large values of $\gamma$). Nevertheless, as the following computation shows, the Wootters concurrence \emph{exactly} matches the X-concurrence in this case.

We first apply the X-concurrence formula to Eq.~(\ref{eq:ss}), which yields
\begin{equation}
C_X(\bm{\rho}_{s,2})=2\max\left[0,\frac{\gamma^2-1}{D_2},\frac{\gamma^2-\sqrt{1+2\gamma^2+4\gamma^4}}{D_2}\right]=\max\left[0,\frac{2(\gamma^2-1)}{D_2}\right]\,.\label{eq:cxrhos}
\end{equation}
Then, in order to compute the Wootters concurrence of $\bm{\rho}_{s,2}$, we start by evaluating
\begin{equation}
\bm{\rho}_{s,2}(\bm{\sigma}_y\otimes\bm{\sigma}_y){[\bm{\rho}_{s,2}]}^\ast(\bm{\sigma}_y\otimes\bm{\sigma}_y)=\frac{1}{D_2^2}\left[
\begin{array}{cccc}
1+4\gamma^4 & 2i\gamma^3 & 2i\gamma^3 & -2\gamma^2\\
4i\gamma^5 & 1-2\gamma^4 & -2\gamma^4 & -2i\gamma^3\\
4i\gamma^5 & -2\gamma^4 & 1-2\gamma^4 & -2i\gamma^3\\
-2\gamma^2-8\gamma^6 & -4i\gamma^5 & -4i\gamma^5 & 1+4\gamma^4
\end{array}
\right]
\end{equation}
whose eigenvalues are
\begin{equation}
\left\{\frac{1+2\gamma^4+2\gamma^2\sqrt{1+\gamma^4}}{D_2^2},\frac{1}{D_2^2},\frac{1}{D_2^2},\frac{1+2\gamma^4-2\gamma^2\sqrt{1+\gamma^4}}{D_2^2}\right\}\,.
\end{equation}
Since the first element in the above set is clearly the largest eigenvalue, one has
\begin{align}
C_W(\bm{\rho}_{s,2})&=\frac{1}{D_2}\max\left[0,\sqrt{1+2\gamma^4+2\gamma^2\sqrt{1+\gamma^4}}-2-\sqrt{1+2\gamma^4-2\gamma^2\sqrt{1+\gamma^4}}\right]\\
&=\max\left[0,\frac{2(\gamma^2-1)}{D_2}\right]\label{eq:crhos}
\end{align}
where, in order to obtain the second row, we used the denesting identities
\begin{equation}
\sqrt{1+2\gamma^4\pm 2\gamma^2\sqrt{1+\gamma^4}}=\sqrt{1+\gamma^4}\pm\gamma^2\,.
\end{equation}
Comparing Eqs.~(\ref{eq:cxrhos}) and~(\ref{eq:crhos}), it is clear that $C_W(\bm{\rho}_{s,2})=C_X(\bm{\rho}_{s,2})$.

Unfortunately, saturation of (\ref{eq:lbCgmeCx}) does not occur for the zero-temperature steady states of the driven Dicke model with $N>2$. For $N=3,4$, for example, density matrices in the computational basis are explicitly presented in Eqs.~(\ref{eq:rhos3cb}) and (\ref{eq:rhos4cb}), respectively. Although Plots~\ref{fig:DIKlowerbound}b,c reveal that these states are GM-entangled for $\gamma\gtrsim 1$, a straightforward computation of theirs X-concurrence yields
\begin{equation}\label{eq:cxrhos3}
C_X(\bm{\rho}_{s,3})=\frac{2}{D_3}\max\left[0,\,6\gamma^3-3\sqrt{\Gamma_1(\Gamma_2+4\gamma^4)},\,2\gamma^3-2\sqrt{\Gamma_1(\Gamma_2+4\gamma^4)}-\sqrt{\Gamma_3(1+12\gamma^4)}\right]=0
\end{equation}
and
\begin{multline}
C_{X}(\bm{\rho}_{s,4})=\frac{2}{D_4}\max\left[0,\,24\gamma^4-3(\Gamma_2+4\gamma^4)-4\sqrt{\Gamma_1\Gamma_3(1+12\gamma^4)},\right.\\
 6\gamma^4-3(\Gamma_2+4\gamma^4)-3\sqrt{\Gamma_1\Gamma_3(1+12\gamma^4)}-\sqrt{\Gamma_4+24\gamma^4\Gamma_6+576\gamma^8},\,\\
\left. 4\gamma^4-2(\Gamma_2+4\gamma^4)-4\sqrt{\Gamma_1\Gamma_3(1+12\gamma^4)}-\sqrt{\Gamma_4+24\gamma^4\Gamma_6+576\gamma^8}\right]=0\,,\label{eq:cxrhos4}
\end{multline}
which shows that $C_X$ is only a trivial lower bound for $C_{GME}$ in these cases.

\section{Density Matrices in the Computational Basis}\label{app:matrixform}
In this appendix we present the density matrices, written in the computational basis, of the $N$-qubit diagonal symmetric states and steady states of the driven Dicke model, for $N=2,3,4$. For ease of visualuzation, we restrict to show the lower triangular entries, replace zeros with dots and boldify the anti-diagonal entries.

\subsection{Diagonal Symmetric States}
\begin{align}\label{eq:rhodss2}
\bm{\rho}_{ds,2}&=\frac{1}{2}\begin{bmatrix}
2p_0 &  &  & \\
\cdot & p_1 &  & \\
\cdot & \bm{p_1} & p_1 & \\
\bm{\cdot} & \cdot & \cdot & 2p_2
\end{bmatrix}
\end{align}

\begin{align}\label{eq:rhodss3}
\bm{\rho}_{ds,3}&=\frac{1}{3}\begin{bmatrix}
3p_0 &  &  &  &  &  &  &  \\
\cdot & p_1 &  &  &  &  &  &  \\
\cdot & p_1 & p_1 &  &  &  &  &  \\
\cdot & \cdot & \cdot & p_2 &  &  &  &  \\
\cdot & p_1 & p_1 & \bm{\cdot} & p_1 &  &  &  \\
\cdot & \cdot & \bm{\cdot} & p_2 & \cdot & p_2 &  &  \\
\cdot & \bm{\cdot} & \cdot & p_2 & \cdot & p_2 & p_2 &  \\
\bm{\cdot} & \cdot & \cdot & \cdot & \cdot & \cdot & \cdot & 3 p_3
\end{bmatrix}
\end{align}

\begin{align}\label{eq:rhodss4}
\bm{\rho}_{ds,4}&=\frac{1}{4}\begin{bmatrix}
4p_0 &  &  &  &  &  &  &  &  &  &  &  &  &  &  & \\
\cdot & p_1 &  &  &  &  &  &  &  &  &  &  &  &  &  & \\
\cdot & p_1 & p_1 &  &  &  &  &  &  &  &  &  &  &  &  & \\
\cdot & \cdot & \cdot & p_2 &  &  &  &  &  &  &  &  &  &  &  & \\
\cdot & p_1 & p_1 & \cdot & p_1 &  &  &  &  &  &  &  &  &  &  & \\
\cdot & \cdot & \cdot & p_2 & \cdot & p_2 &  &  &  &  &  &  &  &  &  & \\
\cdot & \cdot & \cdot & p_2 & \cdot & p_2 & p_2 &  &  &  &  &  &  &  &  & \\
\cdot & \cdot & \cdot & \cdot & \cdot & \cdot & \cdot & p_3 &  &  &  &  &  &  &  & \\
\cdot & p_1 & p_1 & \cdot & p_1 & \cdot & \cdot & \bm{\cdot} & p_1 &  &  &  &  &  &  & \\
\cdot & \cdot & \cdot & p_2 & \cdot & p_2 & \bm{p_2} & \cdot & \cdot & p_2 &  &  &  &  &  & \\
\cdot & \cdot & \cdot & p_2 & \cdot & \bm{p_2} & p_2 & \cdot & \cdot & p_2 & p_2 &  &  &  &  & \\
\cdot & \cdot & \cdot & \cdot & \bm{\cdot} & \cdot & \cdot & p_3 & \cdot & \cdot & \cdot & p_3 &  &  &  & \\
\cdot & \cdot & \cdot & \bm{p_2} & \cdot & p_2 & p_2 & \cdot & \cdot & p_2 & p_2 & \cdot & p_2 &  &  & \\
\cdot & \cdot & \bm{\cdot} & \cdot & \cdot & \cdot & \cdot & p_3 & \cdot & \cdot & \cdot & p_3 & \cdot & p_3 &  & \\
\cdot & \bm{\cdot} & \cdot & \cdot & \cdot & \cdot & \cdot & p_3 & \cdot & \cdot & \cdot & p_3 & \cdot & p_3 & p_3 & \\
\bm{\cdot} & \cdot & \cdot & \cdot & \cdot & \cdot & \cdot & \cdot & \cdot & \cdot & \cdot & \cdot & \cdot & \cdot & \cdot & 4p_4
\end{bmatrix}
\end{align}

\subsection{Steady states of the driven Dicke model}
For brevity, we define the following family (parametrized by $k$) of quadratic functions of $\gamma$:
\begin{equation}
\Gamma_k\coloneq 1+k\gamma^2\,.
\end{equation}
The normalization appearing as a multiplicative factor in front of the matrices forms are given below:
\begin{align}
D_2&\coloneq 4(1+\gamma^2+\gamma^4)\,,\\
D_3&\coloneq 4(2+3\gamma^2+6\gamma^4+9\gamma^6)\,,\\
D_4&\coloneq 16(1+2\gamma^2+6\gamma^4+18\gamma^6+36\gamma^8)\,.
\end{align}

\begin{align}\label{eq:ss}
\bm{\rho}_{s,2}&=\frac{1}{D_2}\begin{bmatrix}
1 &  &  & \\
i\gamma & 1+\gamma^2 &  & \\
i\gamma & \bm{\gamma^2} & 1+\gamma^2 & \\
\bm{-2\gamma^2} & i\gamma (1+2\gamma^2) & i\gamma(1+2\gamma^2) & 1+2\gamma^2+4\gamma^4
\end{bmatrix}
\end{align}

\begin{sideways}
\centering
 \begin{minipage}{\textheight}
\begin{equation}\label{eq:rhos3cb}
\bm{\rho}_{s,3}=\frac{1}{D_3}\left[\begin{array}{cccccccc}
1&&&&&&&\\
\gamma i&\Gamma_1&&&&&&\\
\gamma i&\gamma^2&\Gamma_1&&&&&\\
-2 \gamma^2&\gamma i \Gamma_2&\gamma i \Gamma_2&4 \gamma^4+\Gamma_2&&&&\\
\gamma i&\gamma^2&\gamma^2&\bm{-2 \gamma^3 i}&\Gamma_1&&&\\
-2 \gamma^2&\gamma i \Gamma_2&\bm{2 \gamma^3 i}&\gamma^2\Gamma_4&\gamma i \Gamma_2&4 \gamma^4+\Gamma_2&&\\
-2 \gamma^2&\bm{2 \gamma^3 i}&\gamma i \Gamma_2&\gamma^2\Gamma_4&\gamma i \Gamma_2&\gamma^2\Gamma_4&4 \gamma^4+\Gamma_2&\\
\bm{-6 \gamma^3 i}&-2\gamma^2\Gamma_3&-2\gamma^2\Gamma_3&\gamma i (12\gamma^4+\Gamma_4)&-2\gamma^2\Gamma_3&\gamma i (12\gamma^4+\Gamma_4)&\gamma i (12\gamma^4+\Gamma_4)&(1+12\gamma^4)\Gamma_3
\end{array}\right]
\end{equation}
 \begin{multline}
\bm{\rho}_{s,4}=\frac{1}{D_4}
\left[\begin{smallmatrix}
1&&&&&&&\\
\gamma i&\Gamma_1&&&&&&\\
\gamma i&\gamma^2&\Gamma_1&&&&&\\
-2 \gamma^2&\gamma i\Gamma_2&\gamma i\Gamma_2&4\gamma^4+\Gamma_2&&&&\\
\gamma i&\gamma^2&\gamma^2&-2 \gamma^3 i&\Gamma_1&&&\\
-2 \gamma^2&\gamma i\Gamma_2&2 \gamma^3 i&\gamma^2\Gamma_4&\gamma i\Gamma_2&4\gamma^4+\Gamma_2&&\\
-2 \gamma^2&2 \gamma^3 i&\gamma i\Gamma_2&\gamma^2\Gamma_4&\gamma i\Gamma_2&\gamma^2\Gamma_4&4\gamma^4+\Gamma_2&\\
-6 \gamma^3 i&-2\gamma^2\Gamma_3&-2\gamma^2\Gamma_3&\gamma i(12\gamma^4+\Gamma_4)&-2\gamma^2\Gamma_3&\gamma i(12\gamma^4+\Gamma_4)&\gamma i(12\gamma^4+\Gamma_4)&(1+12\gamma^4)\Gamma_3\\
\gamma i&\gamma^2&\gamma^2&-2 \gamma^3 i&\gamma^2&-2 \gamma^3 i&-2 \gamma^3 i&\bm{-6 \gamma^4}\\
-2 \gamma^2&\gamma i\Gamma_2&2 \gamma^3 i&\gamma^2\Gamma_4&2 \gamma^3 i&\gamma^2\Gamma_4&\bm{4 \gamma^4}&-2i\gamma^3\Gamma_6\\
-2 \gamma^2&2 \gamma^3 i&\gamma i\Gamma_2&\gamma^2\Gamma_4&2 \gamma^3 i&\bm{4 \gamma^4}&\gamma^2\Gamma_4&-2i\gamma^3\Gamma_6\\
-6 \gamma^3 i&-2\gamma^2\Gamma_3&-2\gamma^2\Gamma_3&\gamma i(12\gamma^4+\Gamma_4)&\bm{-6 \gamma^4}&2i\gamma^3\Gamma_6&2i\gamma^3\Gamma_6&\gamma^2(36\gamma^4+\Gamma_8)\\
-2 \gamma^2&2 \gamma^3 i&2 \gamma^3 i&\bm{4 \gamma^4}&\gamma i\Gamma_2&\gamma^2\Gamma_4&\gamma^2\Gamma_4&-2i\gamma^3\Gamma_6\\
-6 \gamma^3 i&-2\gamma^2\Gamma_3&\bm{-6 \gamma^4}&2i\gamma^3\Gamma_6&-2\gamma^2\Gamma_3&\gamma i(12\gamma^4+\Gamma_4)&2i\gamma^3\Gamma_6&\gamma^2(36\gamma^4+\Gamma_8)\\
-6 \gamma^3 i&\bm{-6 \gamma^4}&-2\gamma^2\Gamma_3&2i\gamma^3\Gamma_6&-2\gamma^2\Gamma_3&2i\gamma^3\Gamma_6&\gamma i(12\gamma^4+\Gamma_4)&\gamma^2(36\gamma^4+\Gamma_8)\\
\bm{24 \gamma^4}&-6i\gamma^3\Gamma_4&-6i\gamma^3\Gamma_4&-2\gamma^2(24\gamma^4+\Gamma_6)&-6i\gamma^3\Gamma_4&-2\gamma^2(24\gamma^4+\Gamma_6)&-2\gamma^2(24\gamma^4+\Gamma_6)&i\gamma(\Gamma_6+36\gamma^4\Gamma_4)
  \end{smallmatrix}\right.
\\[10mm]
 \left.\begin{smallmatrix}
&&&&&&&\\
&&&&&&&\\
&&&&&&&\\
&&&&&&&\\
&&&&&&&\\
&&&&&&&\\
&&&&&&&\\
&&&&&&&\\
\Gamma_1&&&&&&&\\
\gamma i\Gamma_2&4\gamma^4+\Gamma_2&&&&&&\\
\gamma i\Gamma_2&\gamma^2\Gamma_4&4\gamma^4+\Gamma_2&&&&&\\
-2\gamma^2\Gamma_3&\gamma i(12\gamma^4+\Gamma_4)&\gamma i(12\gamma^4+\Gamma_4)&(1+12\gamma^4)\Gamma_3&&&&\\
\gamma i\Gamma_2&\gamma^2\Gamma_4&\gamma^2\Gamma_4&-2i\gamma^3\Gamma_6&4\gamma^4+\Gamma_2&&&\\
-2\gamma^2\Gamma_3&\gamma i(12\gamma^4+\Gamma_4)&2i\gamma^3\Gamma_6&\gamma^2(36\gamma^4+\Gamma_8)&\gamma i(12\gamma^4+\Gamma_4)&(1+12\gamma^4)\Gamma_3&&\\
-2\gamma^2\Gamma_3&2i\gamma^3\Gamma_6&\gamma i(12\gamma^4+\Gamma_4)&\gamma^2(36\gamma^4+\Gamma_8)&\gamma i(12\gamma^4+\Gamma_4)&\gamma^2(36\gamma^4+\Gamma_8)&(1+12\gamma^4)\Gamma_3&\\
-6i\gamma^3\Gamma_4&-2\gamma^2(24\gamma^4+\Gamma_6)&-2\gamma^2(24\gamma^4+\Gamma_6)&i\gamma(\Gamma_6+36\gamma^4\Gamma_4)&-2\gamma^2(24\gamma^4+\Gamma_6)&i\gamma(\Gamma_6+36\gamma^4\Gamma_4)&i\gamma(\Gamma_6+36\gamma^4\Gamma_4)&24\gamma^4+(1+144\gamma^6)\Gamma_4
  \end{smallmatrix}\right]\label{eq:rhos4cb}
\end{multline}
\end{minipage}
\end{sideways}


\end{document}